\documentclass[preprint]{elsarticle}

\usepackage{lineno,hyperref}

\usepackage{amsmath} 
\usepackage{xcolor}

\journal{Cognitive Systems Research}

\begin{document}

\begin{frontmatter}

\title{Bucket of deep transfer learning features and classification models for melanoma detection}


\author[mymainaddress]{Mario Manzo\corref{mycorrespondingauthor}}
\cortext[mycorrespondingauthor]{Corresponding author}
\ead{mmanzo@unior.it}

\author[mysecondaryaddress]{Simone Pellino}

\address[mymainaddress]{Information Technology Services,
              University of Naples “L’Orientale”}
\address[mysecondaryaddress]{MIUR}

\begin{abstract}
Malignant melanoma is the deadliest form of skin cancer and, in recent years, is rapidly growing in terms of the incidence worldwide rate. The most effective approach to targeted treatment is early diagnosis. Deep learning algorithms, specifically convolutional neural networks, represent a methodology for the image analysis and representation. They optimize the features design task, essential for an automatic approach on different types of images, including medical. In this paper, we adopted pretrained deep convolutional neural networks architectures for the image representation with purpose to predict skin lesion melanoma. Firstly, we applied a transfer learning approach to extract image features. Secondly, we adopted the transferred learning features inside an ensemble classification context. Specifically, the framework trains individual classifiers on balanced subspaces and combines the provided predictions through statistical measures. Experimental phase on datasets of skin lesion images is performed and results obtained show the effectiveness of the proposed approach with respect to state-of-the-art competitors.
\end{abstract}


\end{frontmatter}


\section{Introduction}
\label{intro}

Among the types of malignant cancer, melanoma is the deadliest form of skin cancer and its incidence rate is growing rapidly around the world. Early diagnosis is particularly important since melanoma can be cured with a simple excision. In the majority, due to the similarity of the various skin lesions (melanoma and not-melanoma) \cite{codella2015deep}, the visual analysis could be unsuitable and would lead to a wrong diagnosis. In this regard, image processing and artificial intelligence tools can provide a fundamental aid to a step of automatic classification \cite{mishra2016overview}. Further improvement in diagnosis is provided by dermoscopy technique \cite{binder1995epiluminescence}. Dermoscopy technique can be applied to the skin, in order to capture illuminated and magnified images of the skin lesion in a non invasive way to highlight areas containing spots. Furthermore, the visual effect of the deeper skin layer can be improved if the skin surface reflection is removed. Anyhow, classification of melanoma dermoscopy images is a difficult task for different issues. First, the degree of similarity between melanoma and not-melanoma lesions. Second, the segmentation, and, therefore, the identification of the affected area is very complicated because of the variations in terms of texture, size, color, shape and location. The last issue and not the least, is the additional skin conditions such as hair, veins or variations due to image capturing. To this end, many solutions have been provided to improve the task. For example, low-level hand-crafted features \cite{barata2018survey} are adopted to discriminate non-melanoma and melanoma lesions. In some cases, this type of features are unable to discriminate clearly, leading to results that are sometimes not very relevant \cite{celebi2007methodological}. Differently, segmentation is adopted to isolate the foreground elements from the background ones \cite{tommasi2006melanoma}. Consequently, the segmentation includes low-level features with a low representational power that provides unsatisfactory results \cite{pathan2018methodological}. In recent years, deep learning has become an effective solution for the extraction of significant features on large data. In particular, the diffusion of deep neural networks, applied to the image classification task, is connected to various factors such as the availability of software in terms of open source license, the constant growth of hardware power and the availability of large datasets \cite{Alexnet}. Deep learning has proven effective for the management, analysis, representation and classification of medical images \cite{ronneberger2015u}. Specifically, for the treatment of melanoma, deep neural networks were adopted both in segmentation and classification phases \cite{yu2016automated}. However, the high variation of the types of melanoma and the imbalance of the data have a decisive impact on performance \cite{shie2015transfer}, hindering the generalization of the model and leading to over-fitting \cite{shin2016deep}. In order to overcome the aforementioned issues, in this paper, we introduce a novel framework based on transfer deep learning and ensemble classification for melanoma detection. It works based on three integrated stages. A first, which performs image preprocessing operations. A second, which extracts features using transfer deep learning. A third, including a layer of ensemble learning, in which different classification algorithms and features extracted are combined with the aim of making the best decision (melanoma/not-melanoma). 
Our approach provides the following main contributions:
\begin{itemize}
\item A deep and ensemble learning based framework, to simultaneously address inter-class variation and class imbalance for the task of melanoma classification.
\item A framework that in the classification phase, at the same time, creates multiple image representation models, based on features extracted with deep transfer learning.
\item The demonstration of how the choice of multiple features can enrich image representation by leading a lesion assessment like a skilled dermatologist.
\item Some experimental greater improvements over existing methods on different state of art datasets about melanoma detection task.
\end{itemize}

The paper is structured as follows. Section \ref{relatedwork} provides an overview of state of art about melanoma classification approaches. Section \ref{MM} describes in detail proposed framework. Section \ref{res} provides a wide experimental phase, while section \ref{conc} concludes the paper.


\section{Related work}
\label{relatedwork}
In this section, we briefly analyze the most important approaches of skin lesions recognition literature. In this field are included numerous works that address the issue according to different aspects. Some works offer an important contribution about image representation, by implementing segmentation algorithms or new descriptors. Instead, others implement complex mechanisms of learning and classification. 

In \cite{Colorpigmented}  a novel boundary descriptor based on the color variation of the skin lesion input images, achieved with standard cameras is introduced. Furthermore, in order to reach higher performance, a set of textural and morphological features is added. Multilayer perceptron neural network as classifier is adopted.

In \cite{melanomadecision} authors propose a complex framework that implements an illumination correction and features extraction on skin image lesions acquired using normal consumer-grade cameras.  Applying a multi-stage illumination improvement algorithm and defining a set of high-level intuitive features (HLIF), that quantifies the level of asymmetry and border irregularity about a lesion, the proposed model can be used to classify accurate skin lesion diagnoses.

While in \cite{bibis} authors, to properly evaluate contents of the concave contours, introduce a
novel border descriptor named boundary intersection-based signature (BIBS). Shape signature is a one-dimensional illustration of shape border and cannot contribute to a proper description for concave borders that have more than one intersection points. For this reason, BIBS analyzes boundary contents of shape especially shapes with concave contours. Support vector machine (SVM) for classification process is adopted. 

Another descriptor for the individualization of skin lesions is named high-level intuitive features (HLIFs) \cite{hlifs}. HLIFs is created to simulate a model of human-observable characteristics. It captures specific characteristics that are significant to the given application: Color Asymmetry, analyzing and clustering pixels colors, Structural Asymmetry, applying the Fourier descriptors of the shape, Border Irregularity, using morphological opening and closing, Color characteristics, transforming the image to a perceptually uniform color space, building color-spatial representations that model the color information for a patch of pixels, clustering the patch representations into $k$ color clusters, quantifying the variance found using the original lesion and the $k$ representative colors. 

A texture analysis method of Local Binary Patterns (LBP) and Block Difference of Inverse Probabilities is proposed in \cite{Textureanalysis}. A comparison is provided with classification results obtained by taking the raw pixel intensity values as input. Classification stage is achieved generating an automated model obtained by both Convolutional Neural Networks (CNN) and SVM.

In \cite{MED-NODE} authors propose a system that automatically extracts the lesion regions, using a non-dermoscopic digital images, and then computes color and texture descriptors. Extracted features are adopted for automatic prediction step. The classification is managed using a majority vote of all predictions.

In \cite{Nasr} non-dermoscopic clinical images to assist a dermatologist in early diagnosis of melanoma skin cancer are adopted. Images are preprocessed in order to reduce artifacts like noise effects. Subsequently, images are analyzed through a pretrained CNN which is a member of deep learning models. CNN are trained by large number of training samples in order to distinguish between melanoma and benign cases.

In \cite{ensemble} Predict-Evaluate-Correct K-fold (PECK) algorithm is presented. Algorithm works by merging deep CNNs with SVM and random forest classifiers to achieve an introspective learning method. In addition, authors provides a novel segmentation algorithm, named Synthesis and Convergence of Intermediate Decaying Omnigradients (SCIDOG), to accurately detect lesion contours in non-dermoscopic images, even in the presence of significant noise, hair, and fuzzy lesion boundaries. 

In \cite{border-line} authors propose a novel solution to improve melanoma classification by defining a new feature that exploits the border-line characteristics of the lesion segmentation mask combining gradients with LBP.
These border-line features are used together with the conventional ones and lead to higher accuracy in classification stage.

In \cite{DeepPCA} an objective features extraction function for CNN is proposed. The goal is to acquire the variation separability as opposed to the categorical cross entropy which maximizes according to the target labels. The deep representative features increase the variance between the images making it more discriminative. Also, the idea is to build a CNN and perform principal component analysis (PCA) during the train phase.

In \cite{MelanomaSegmentation} a deep learning computer aided diagnosis system for automatic segmentation and classification of melanoma lesions is proposed. The system extracts CNN and statistical and contrast location features on the results of raw image segmentation. The combined features are utilized to obtain the final classification of melanoma, malignant or benign. 

In \cite{Automatic-Detection-of-Melanoma} authors propose an efficient algorithm for prescreening of pigmented skin lesions for malignancy using general-purpose digital cameras. The proposed method enhances borders and extracts a broad set of dermatologically important features. These discriminative features allow classification of lesions into two groups of melanoma and benign.

In \cite{accessible} a skin lesion detection system optimized to run entirely on the resource constrained smartphone is described. The system combines a lightweight method for skin detection with a hierarchical segmentation approach including two fast segmentation algorithms and proposes novel features to characterize a skin lesion. Furthermore, the system implements an improved features selection algorithm to determine a small set of discriminative features adopted by the final lightweight system.


\section{Materials and Methods}
\label{MM}
In the section we describe the proposed framework which includes two well known methodologies: deep neural network and ensemble learning. The main idea is to combine algorithms of features extraction and classification. The result is a set of competitive models providing a range of confidential decisions useful for making choices during classification. The framework is composed of three level. A first, which performs preprocessing operations such as image resize and data balancing. A second, of transfer learning, which extracts features using deep neural networks. A third level, of ensemble learning, in which different classification algorithms (SVM \cite{svm}, Logistic Label Propagation (LLP) \cite{kobayashi2012logistic}, KNN \cite{knn}) and features extracted are combined with the aim of making the best decision. Adopted classifiers are trained and tested through a bootstrapping policy. Finally, the framework iterates through a predetermined number of times in a supervised learning context.



\subsection{Data balancing}
\label{databalancing}
Melanoma lesion analysis and classification is connected with accurate segmentation with purpose to isolate areas of the image containing information of interest. Moreover, the wide variety of skin lesions and the unpredictable obstructions on the skin make traditional segmentation an ineffective tool, especially for non-dermoscopic
images. Furthermore, the problem of imbalance, present in many datasets, makes the classification difficult to address, especially when the samples of the minority class are very underrepresented. In the case under consideration, to compensate the strong imbalance between the two classes, a balancing phase was performed. The goal is to isolate segments of the image that could contain melanoma. In particular, the resampling of the minority class is performed by adding images altered through the application of K-Means color segmentation algorithm \cite{likas2003global}. The application of segmentation algorithms for image augmentation \cite{shorten2019survey}, and consequently to provide a balancing between classes, represented a good compromise for this stage of the pipeline.

\subsection{Image resize}
\label{resize}

Images to be processed have been resized based on the dimension, related to the input layer, claimed by the deep neural networks (details can be found in table \ref{nets} column 5). Many of the networks require this type of step but it does not alter the image information content in any way. This normalization step is essential because images of different or large dimensions cannot be processed for the features extraction stage.

\subsection{Transfer learning and features extraction}
\label{Bucket of nets}

The transfer learning approach has been chosen for features extraction purpose. Commonly, pretrained network is adopted as starting point to learn a new task. It is the easiest and fastest way to exploit the representational power of pretrained deep networks. It is usually much faster and easier to tune a network with transfer learning
than training a new network from scratch with randomly initialized weights. We have selected deep learning architectures for image classification based on their structure and performance skills. The goal is to extract features from images through neural networks by redesign their structures in the final layer according to the needs of the addressed task (two outgoing classes: melanoma and not-melanoma). The features extraction is performed through a chosen layer (different for each network and specified in the table \ref{nets}), placed in the final part of the structure. The image will be encoded through a vector of real numbers produced by consecutive convolution steps, from the input layer to the layer chosen for the representation. Below a description of the adopted networks is reported.

Alexnet \cite{Alexnet} consists of 5 convolutional layers and 3 fully connected layers. It includes the non-saturating ReLU activation function, better then tanh and sigmoid during training phase. For features extraction, we have chosen fully connected 7 (fc7) layer composed of 4096 neurons.

Googlenet \cite{googlenet} is composed of 22 layers deep. The network is inspired by LeNet \cite{lecun1989backpropagation} but implemented a novel element which is dubbed an inception module. This module is based on several very small convolutions in order to drastically reduce the number of parameters. Their architecture reduced the number of parameters from 60 million (AlexNet) to 4 million. Furthermore, it includes batch normalization, image distortions and Root Mean Square Propagation algorithm. For features extraction, we have chosen global average pooling (pool5-7x7\_s1) layer composed of 1024 neurons.

Resnet18 and  Resnet50 \cite{Resnet} are inspired by pyramidal cells contained in the cerebral cortex. They use particular skip connections or shortcuts to jump over some layers. They are composed of 18 and 50 layers deep, which with the help of a technique known as skip connection has paved the way for residual networks. For features extraction, we have chosen two global average pooling (pool5 and avg-pool) layers composed of 512 and 2048 neurons respectively.

\begin{table}[!ht]
\centering \caption{Description of adopted pretrained network.}
\footnotesize
\begin{tabular}{|l|l|l|l|l|l|}
\hline
  Network & Depth & Size (MB) & Parameters (Millions) & Input Size & Features Layer\\ \hline
  Alexnet & 8 & 227 & 61 & 227 $\times$ 227 & fc7\\ \hline
  Googlenet & 8 & 27 & 7 & 224 $\times$ 224 & pool5-7x7\_s1\\ \hline
  Resnet18 & 18 & 44 & 11.7 & 224 $\times$ 224 & pool5\\ \hline
  Resnet50 & 50 & 96 & 25.6 & 224 $\times$ 224 & avg\_pool\\ \hline
\end{tabular}
\label{nets}
\end{table}














\subsection{Network design}
\label{netmod}

The adopted networks  have been adapted to the melanoma classification problem. Originally, they have been trained on the Imagenet dataset \cite{deng2009imagenet}, composed of a million images and classified into 1000 classes. The result is a rich features representation for a wide range of images. The network processes an image and provides a label along with probabilities for each of the classes. Commonly, the first layer of the network is the image input layer. This requires input images with 3 color channels. Just after, convolutional layers work to extract image features in which the last learnable layer and the final classification layer adopt to classify the input image. In order to make suitable the pretrained network to classify new images, the two last layers with new layers are replaced. In many cases, the last layer, including learnable weights, is a fully connected layer. This is replaced with a new fully connected layer related to the number of outputs equal to the number of classes of new data. Moreover, to speedup the learning in the new layer respect to transferred layers, it is recommended to increase the learning rate factors. As an optional choice, the weights of earlier layers can be frozen by setting the related learning rate to zero. This setting produces a failure of update of the weights, during the training, and a consequent lowering of the execution time as the gradients of the related layers must not be calculated. This aspect is very interesting to avoid overfitting in the case of small datasets.





\subsection{Ensemble Learning}
\label{ense}


The contribution of different transfer learning features and classifiers can be mixed in an ensemble context. Considering the set of images, with cardinality $k$, belonging to $x$ classes, to be classified

\begin{equation}
Imgs=\{i_{1},i_{2},\ldots,i_{k}\}    
\end{equation}

each element of the set will be treated with the procedure below. Let's consider the set $C$ composed of $n$ classifiers

\begin{equation}
C=\{\beta_{1},\beta_{2},\ldots,\beta_{n}\}    
\end{equation}
and set $F$ composed of $m$ vectors of transferred learning features 

\begin{equation}
\label{feats}
F=\{\Theta_{1},\Theta_{2},....\Theta_{m}\}    
\end{equation}

the goal is the combination each element of the set $C$ with the elements of the set $F$. The set of combinations can be defined as $CF$


\[
CF = \begin{bmatrix} 
    \beta_{1}\Theta_{1} & \dots & \beta_{1}\Theta_{m} \\
    \vdots & \ddots & \\
    \beta_{n}\Theta_{1} &        & \beta_{n}\Theta_{m} 
    \end{bmatrix}
\]

each combination provides a decision $i \in I \{-1,1\}$, where $1$ stands for melanoma and $-1$ for not-melanoma, related to image of the set $Imgs$. The set of decisions $D$ can be defined as follows


\[
D = \begin{bmatrix} 
    d_{\beta_{1}\Theta_{1}} & \dots & d_{\beta_{1}\Theta_{m}} \\
    \vdots & \ddots & \\
    d_{\beta_{n}\Theta_{1}} &        & d_{\beta_{n}\Theta_{m}} 
    \end{bmatrix}
\]

Each $d_{\beta_{i}\Theta_{j}}$ value represents a decision based on the combination of sets $C$ and $F$. In addition, the set of scores S can be defined as follows

\[
S = \begin{bmatrix} 
    P(i|x)_{d_{\beta_{1}\Theta_{1}}} & \dots & P(i|x)_{d_{\beta_{1}\Theta_{m}}} \\
    \vdots & \ddots & \\
    P(i|x)_{d_{\beta_{n}\Theta_{1}}} &        & P(i|x)_{d_{\beta_{n}\Theta_{m}}} 
    \end{bmatrix}
\]

a score value, $s \in S \{0,\dots,1\}$, is associated with each decision $d$ and represents the posterior probability $P(i|x)$ that an image $i$ belongs to class $x$.  At this point, let's introduce the concept of mode, defined as the value which is repeatedly occurred in a given set

\begin{equation}
mode=l+\left(\frac{f_1-f_0}{2f_1-f_0-f_2}\right)
\times h
\end{equation}

where $l$ is the lower limit of the modal class, $h$ is the size of the class interval, $f_1$ is the frequency of the modal class, $f_0$ is the frequency of the class which precedes the modal class and $f_2$ is the frequency of the class which successes the modal class. The columns of matrix $D$ are analyzed with the mode, in order to obtain the values of the most frequent decisions. This step is carried out in order to verify the best response of the different classifiers, contained in the set $C$, which adopt the same type of features. Moreover, the $mode$ provides two indications. The most frequent value and its occurrences (indices). For each most frequent occurrence, modal value, the corresponding score of the matrix $S$ is extracted. In this regard, a new vector is generated

\begin{equation}
DS=\{ds_{P(i|x)_{d_{\beta_{1,\dots,n}\Theta_{1}}}}, \ldots, ds_{P(i|x)_{d_{\beta_{1,\dots,n}\Theta_{m}}}} \},\end{equation}

where each element $ds$ contains the average of the scores that have a higher frequency, extracted through the $mode$, in the related column of the matrix $D$. Also, the modal value of each column of the matrix $D$ is stored in the vector $DM$

\begin{equation}
DM=\{dm_{d_{\beta_{1,\dots,n}\Theta_{1}}},\dots, dm_{d_{\beta_{1,\dots,n}\Theta_{m}}}\},\end{equation}

the final decision will consist in the selection of the element of the vector $DM$ with the same position of the maximum score value of the vector $DS$. This last step verifies the best prediction based on the different features adopted, essentially the best features suitable for the classification of the image.





\subsection{Train and test strategy: Bootstrapping}
\label{traintest}

Bootstrapping is a statistical technique which consists in creating samples of size $B$, named bootstrap samples, from a dataset of size $N$.
The bootstrap samples are random inserted with replacement on the dataset. This strategy has important statistical properties. First, subsets can be considered as directly extracted from the original distribution, independently of each others, containing representative and independent samples, almost independent and identically distributed (idd). Two considerations must be made in order to validate the hypotheses. First, the $N$ dimension of the original dataset should be large enough to detect the underlying distribution. Sampling the original data is a good approximation of real distribution (representativeness). Second, the $N$ dimension of the dataset should be better than the $B$ dimension of the bootstrap samples so that the samples are not too correlated (independence). Commonly, considering the samples to be truly independent means requiring too much data compared to the amount actually available. This strategy can be adopted to generate several bootstrap samples that can be considered nearly representative and almost independent (almost iid samples). In the proposed framework, bootstrapping is applied to set $F$ (equation \ref{feats}) in order to perform the training and testing stages of classifiers. This strategy seemed suitable for the problem faced in order to create a competitive environment capable of providing the best performance.




\section{Experimental results}
\label{res}

This section describes the experiments performed on public datasets. In order to produce compliant performance, the settings included in well-known melanoma classification methods, in which the main critical issue concerns the features extraction for image representation, are adopted.

\subsection{Datasets}
\label{dataset}

First adopted dataset is MED-NODE\footnote{http://www.cs.rug.nl/~imaging/databases/melanoma\_naevi/}. It was created by the Department of Dermatology
of the University Medical Center Groningen (UMCG). The
dataset was initially used to train the MED-NODE computer
assisted melanoma detection system \cite{giotis2015med}. It is composed of 170 non-dermoscopic images, where 70 are melanoma and 100 are nevi. The image dimensions vary greatly, ranging from $201 \times 257$ to $3177 \times 1333$ pixels.

Second adopted dataset, Skin-lesion (from now), is described in \cite{amelard2014high}. It is composed of $206$ images of skin lesion, which were obtained
using standard consumer-grade cameras in varying and
unconstrained environmental conditions. These images were
extracted from the online public databases Dermatology Information
System\footnote{http://www.dermis.net} and DermQuest\footnote{http://www.dermquest.com}. Of these images,
$119$ are melanomas, and $87$ are not-melanoma. Each image
contains a single lesion of interest.

\subsection{Settings}
\label{settings}

The framework consists of different modules written in Matlab language. Moreover, we applied pretrained networks available which are included in the ImageNet Large-Scale Visual Recognition Challenge (ILSVRC) \cite{russakovsky2015imagenet}. Among all the computational stages, the features extraction process, described in section \ref{Bucket of nets}, was certainly the most expensive. As is certainly known, the networks are composed of fully connected layers that make the structure extremely dense and complex. This aspect certainly increases the computational load. Alexnet, Googlenet, Resnet50 are adopted to extract features on MED-NODE dataset. Differently, Resnet50 and Resnet18 are adopted for Skin-lesion dataset. In the table \ref{nets}, some important details related to the layers chosen for features extraction are shown. Networks were trained by setting the mini batch size to $5$, the maximum epochs to $10$, the initial learning rate to $3 \cdot 10^{-4}$ and the optimizer is stochastic gradient descent with momentum (SGDM) algorithm. For both experimental procedures, in order to train the classifiers, $80\%$ and $20\%$ of images are included in train and test set respectively, for a number of iteration equal to $10$. Table \ref{Algorithms} enumerates classification algorithms included in the framework and related settings (some algorithms appear more times with different configurations).

\begin{table}[!ht]
\centering
\caption{Classification algorithms and related settings.}
\tiny
\begin{tabular}{c c }
 &      \\

\hline \textbf{Algorithms}   & \textbf{Setting} \\

\hline  SVM \cite{svm}         & KernelFunction:polynomial, KernelScale: auto                     \\

\hline SVM \cite{svm}         & KernelFunction: gaussian, KernelScale: auto                     \\

\hline LLP \cite{kobayashi2012logistic}        & KernelFunction: rbf, Regularization parameter: 1, init:0, maxiter: 1000           \\

 \hline KNN \cite{knn}         & NumNeighbors: 3, Distance: spearman           \\
 
  \hline KNN \cite{knn}         &  NumNeighbors: 4, Distance: correlation            \\

\hline
  &                                                                      
\end{tabular}
\label{Algorithms}
\end{table}

\subsection{Discussion}
\label{discussion}

The tables \ref{comparisonMed} and \ref{comparisonSkin} describe the comparison with existing skin cancer classification methods (we referred with the results which appear in the corresponding papers). The provided performance can be considered satisfactory compared to competitors. In terms of accuracy, although it provides a rough measurement, we have provided the best result for MED-NODE and the second for Skin-lesion (only surpassed by BIBS). Differently, PPV and NPV give good indications on the classification ability. TPR, a measure that provides greater confidence about addressed problem, is very high for both datasets. Otherwise, TNR, which also provides a high degree of sensitivity related to the absence of tumors within the image, is the best value for both datasets. Regarding the remaining measures, $F_{1}^{p}$, $F_{1}^{n}$ and MCC, considerable values were obtained but, unfortunately, not available for all competitors. We can certainly attribute the satisfactory performance to two main aspects. First, the deep learning features, which even if abstract, are able to best represent the images. Furthermore, the framework provides multiple representation models that certainly constitute a different starting point than a standard approach, in which a single representation is provided. This aspect is relevant for improving performance. Not negligible issue, the normalization of the image size, with respect to the request of the first layer of the neural network, before the features extraction phase, does not produce a performance degradation. In other cases, normalization causes loss of quality of the image content and a consequent degradation of details. Otherwise, the weak point is the computational load even if pretrained networks include layers with already tuned weights. Surely, the time required for training is long but less than a network created from scratch. Second, the classification scheme, which provides multiple choices in decision making. In fact, at each iteration, the framework chooses which classifier is suitable for recognizing melanoma in the images included in the proposed set. Certainly, this approach is more computationally expensive but produces better results than a single classifier.

Moreover, the table \ref{Metric} shows the metrics adopted for the performance evaluation,
in order to provide a uniform comparison with algorithms working on the same task.

\begin{table}[!ht]
\centering
\caption{Evalutation metrics adopted during relevance feedback stage.}
\tiny
\begin{tabular}{c c}
 &      \\

\hline \textbf{Metric}   & \textbf{Equation}  \\

\hline True Positive Rate         & $TPR = \frac{TP}{TP + FN}$                     \\

\hline  True Negative Rate      & $TNR = \frac{TN}{TN + FP}$                    \\

\hline Positive Predictive Value         & $PPV = \frac{TP}{TP + FP}$                     \\

\hline Negative Predictive Value         & $NPV = \frac{TN}{TN + FN}$                     \\

\hline Accuracy         & $ACC = \frac{TP+FN}{TP + FP + TN + FN}$                     \\

\hline $F_{1}$-Score(Positive)         & $F_{1}^{P} = \frac{2 \cdot PPV \cdot TPR}{PPV + TPR}$                     \\

\hline $F_{1}$-Score(Negative)         & $F_{1}^{N} = \frac{2 \cdot NPV \cdot TNR}{NPV + TNR}$                     \\

\hline Matthew's Correlation Coefficient         & $MCC = \frac{TP \cdot TN - FP\cdot FN}{

\sqrt{(TP + FP) \cdot (TP + FN ) \cdot (TN + FP) \cdot (TN + FN )}}$                      \\

\hline
 &                                                                      
\end{tabular}
\label{Metric}
\end{table}

Looking carefully at the table, it is important to focus on the meaning of the individual measures with reference to melanoma detection. The True Positive rate, also known as Sensitivity, concerns the portion of positives melanoma images that are correctly identified. This provide important information because highlights the skill to identify images containing skin lesions and contributes to increase the degree of robustness of result. The same concept is true for the True Negative rate, also known as Specificity, which instead measures the portion of negatives, not containing skin lesions, that have been correctly identified. The Positive and Negative Predictive values, also known as Precision and Recall respectively, are probabilistic measures that indicate whether an image with a positive or negative melanoma test may or may not have a skin lesion. In essence, Recall expresses the ability to find all relevant instances in the dataset, Precision expresses the proportion of instances that the framework claims to be relevant were actually relevant. Accuracy, a well-known performance measure, is the proportion of true results among the total number of cases examined. In our case provides an overall analysis, certainly a rough measurement compared to the previous ones, about the skill of a classifier to distinguish a skin lesion from an image without lesions. $F_{1}-Score$ measure combines the Precision and Recall of the model, as the harmonic mean, in order to find an optimal blend. The choice of the harmonic mean instead of a simple mean concerns the possibility of eliminating extreme values. Finally, Matthew's correlation coefficient is another overall well-known quality measure. It takes into account True/False Positives/Negatives values and is generally regarded as a balanced measure which can be adopted even if the classes are of very different sizes.





 
\begin{table}[!ht]
\centering \caption{Experimental results on MED-NODE dataset.}
\tiny
\begin{tabular}{lllllllllll}
 &  &                                       &  &  &  &  &  &  &  &      
                    \\
 \hline  \multicolumn{3}{l}{Method}          & \multicolumn{8}{l}{ TPR  \quad TNR \quad PPV \quad NPV \quad  ACC  \quad $F_{1}^{p}$  \qquad $F_{1}^{n}$ \quad  MCC}   \\
\hline \multicolumn{3}{l}{MED-NODE annoted \cite{MED-NODE}}  & \multicolumn{8}{l}{0.78 \quad 0.59 \qquad 0.56 \quad  0.80 \qquad 0.66 \quad 0.65    \quad 0.68    \quad 0.36}  \\
\hline\multicolumn{3}{l}{Spotmole \cite{spotmole}}                & \multicolumn{8}{l}{0.82  \quad      0.57  \qquad    0.56 \quad    0.83 \qquad    0.67 \quad     0.67 \quad      0.68   \quad       0.39}  \\
\hline\multicolumn{3}{l}{Barhoumi and Zagrouba \cite{Barhoumi-Zagrouba}}   & \multicolumn{8}{l}{0.46  \quad      0.87  \qquad    0.70 \quad    0.71 \qquad    0.70 \quad     0.56 \quad     0.78  \quad        0.37}  \\
\hline\multicolumn{3}{l}{MED-NODE color \cite{MED-NODE}}          & \multicolumn{8}{l}{0.74  \quad       0.72 \qquad     0.64 \quad    0.81  \qquad   0.73 \quad      0.69 \quad      0.76   \quad       0.45}  \\
\hline\multicolumn{3}{l}{MED-NODE texture \cite{MED-NODE}}        & \multicolumn{8}{l}{0.62   \quad     0.85 \qquad     0.74  \quad   0.77 \qquad    0.76 \quad     0.67 \quad      0.81  \quad         0.49}  \\
\hline\multicolumn{3}{l}{Jafari et al. \cite{Automatic-Detection-of-Melanoma}}           & \multicolumn{8}{l}{0.90    \quad    0.72 \qquad     0.70 \quad    0.91 \qquad    0.79  \quad     0.79 \quad      0.80   \quad        0.61}  \\
\hline\multicolumn{3}{l}{MED-NODE combined \cite{MED-NODE}}       & \multicolumn{8}{l}{0.80    \quad    0.81 \qquad     0.74 \quad    0.86 \qquad    0.81 \quad      0.77 \quad      0.83  \quad        0.61}  \\
\hline\multicolumn{3}{l}{Nasr Esfahani et al. \cite{Nasr}}    & \multicolumn{8}{l}{0.81    \quad    0.80 \qquad     0.75 \quad    0.86 \qquad    0.81 \quad     0.78 \quad      0.83 \quad         0.61}  \\

\hline\multicolumn{3}{l}{Benjamin Albert \cite{ensemble}}           & \multicolumn{8}{l}{0.89    \quad    0.93  \qquad    0.92 \quad    0.93 \qquad    0.91 \quad      0.89 \quad      0.92 \quad          0.83}  \\
\hline\multicolumn{3}{l}{Pereira et\cite{border-line}ght/svm-smo/f23-32}           & \multicolumn{8}{l}{0.45    \quad    0.92  \qquad ~~~    $-$ ~  \quad    $-$  \qquad 0.73 ~    \quad      ~  $-$  \quad ~      $-$  \quad ~           $-$ }  \\

\hline\multicolumn{3}{l}{Pereira et \cite{border-line}ght/svm-smo/f1-32}           & \multicolumn{8}{l}{0.56    \quad    0.86  \qquad  ~~~  $-$ \quad  ~   $-$  \qquad 0.74  ~  \quad    ~   $-$  \quad    ~   $-$  \quad ~          $-$ }  \\

\hline\multicolumn{3}{l}{Pereira et al. \cite{border-line}lbpc/svm-smo/f23-32}           & \multicolumn{8}{l}{0.49    \quad    0.93  \qquad    ~~~  $-$\quad ~~   $-$  \qquad  0.75 ~  \quad ~     $-$  \quad ~     $-$  \quad ~         $-$ }  \\

\hline\multicolumn{3}{l}{Pereira et al. \cite{border-line}lbpc/svm-smo/f1-32}           & \multicolumn{8}{l}{0.58    \quad    0.91  \qquad    ~~~  $-$\quad ~~   $-$  \qquad 0.78    ~  \quad ~     $-$  \quad ~     $-$  \quad ~         $-$ }  \\

\hline\multicolumn{3}{l}{Pereira et al. \cite{border-line}ght/svm-sda/f23-32}           & \multicolumn{8}{l}{0.66    \quad    0.83  \qquad  ~~~  $-$ \quad     ~~   $-$  \qquad 0.76    ~  \quad ~     $-$  \quad ~     $-$  \quad ~         $-$}  \\

\hline\multicolumn{3}{l}{Pereira et al. \cite{border-line}ght/svm-sda/f1-32}           & \multicolumn{8}{l}{0.66    \quad    0.86  \qquad    ~~~  $-$ \quad    ~~   $-$  \qquad 0.78    ~  \quad ~     $-$  \quad ~     $-$  \quad ~         $-$}  \\

\hline\multicolumn{3}{l}{Pereira et al. \cite{border-line}lbpc/svm-isda/f23-32}           & \multicolumn{8}{l}{0.69    \quad    0.83  \qquad    ~~~  $-$ \quad    ~~   $-$  \qquad 0.77   ~  \quad ~     $-$  \quad ~     $-$  \quad ~         $-$ }  \\

\hline\multicolumn{3}{l}{Pereira et al. \cite{border-line}lbpc/svm-isda/f1-32}           & \multicolumn{8}{l}{0.65    \quad    0.88  \qquad    ~~~  $-$ \quad     ~~   $-$   \qquad 0.79       ~  \quad ~     $-$  \quad ~     $-$  \quad ~         $-$ }  \\

\hline\multicolumn{3}{l}{Pereira et al. \cite{border-line}ght/ffn/f23-32}           & \multicolumn{8}{l}{0.63    \quad    0.84  \qquad    ~~~  $-$ \quad    ~~   $-$ \qquad 0.76    ~  \quad ~     $-$  \quad ~     $-$  \quad ~         $-$ }  \\

\hline\multicolumn{3}{l}{Pereira et al. \cite{border-line}ght/ffn/f1-32}           & \multicolumn{8}{l}{0.63    \quad    0.84  \qquad    ~~~  $-$ \quad   ~~   $-$  \qquad 0.76   ~  \quad ~     $-$  \quad ~     $-$  \quad ~         $-$ }  \\

\hline\multicolumn{3}{l}{Pereira et al. \cite{border-line}lbpc/ffn/f23-32}           & \multicolumn{8}{l}{0.64    \quad    0.83  \qquad    ~~~  $-$  \quad     ~~   $-$   \qquad 0.75    ~  \quad ~     $-$  \quad ~     $-$  \quad ~         $-$ }  \\

\hline\multicolumn{3}{l}{Pereira et al. \cite{border-line}lbpc/ffn/f1-32}           & \multicolumn{8}{l}{0.66    \quad    0.86  \qquad    ~~~  $-$ \quad     ~~   $-$   \qquad 0.77    ~  \quad ~     $-$  \quad ~     $-$  \quad ~         $-$}  \\

\hline\multicolumn{3}{l}{Sultana et al. \cite{DeepPCA}}           & \multicolumn{8}{l}{0.73    \quad    0.86  \qquad    0.77 ~~~    0.83  \qquad 0.81 ~    \quad      ~     $-$  \quad ~     $-$  \quad ~         $-$}  \\

\hline\multicolumn{3}{l}{Ge, Yunhao and Liet al. \cite{MelanomaSegmentation}} & \multicolumn{8}{l}{0.94    \quad    0.93  \qquad     ~~~  $-$ \quad    ~~   $-$   \qquad 0.92  ~  \quad    ~     $-$  \quad ~     $-$  \quad ~         $-$ }  \\


\hline\multicolumn{3}{l}{Mandal et al.\cite{deeepresidual} Case 1} & \multicolumn{8}{l}{  0.61  \quad 0.65     \qquad 0.74   \quad 0.87     \qquad 0.65 ~   \quad      ~     $-$  \quad ~     $-$  \quad ~         $-$ }  \\

\hline\multicolumn{3}{l}{Mandal et al.\cite{deeepresidual} Case 2} & \multicolumn{8}{l}{  0.80  \quad 0.73     \qquad 0.74   \quad 0.87     \qquad 0.71 ~   \quad       ~     $-$  \quad ~     $-$  \quad ~         $-$ }  \\

\hline\multicolumn{3}{l}{Mandal et al.\cite{deeepresidual} Case 3} & \multicolumn{8}{l}{  0.84  \quad 0.66     \qquad 0.68   \quad 0.86     \qquad 0.71 ~  \quad      ~     $-$  \quad ~     $-$  \quad ~         $-$  }  \\

\hline\multicolumn{3}{l}{Jafari et al. \cite{set-of-descriptor}} & \multicolumn{8}{l}{0.82    \quad    0.71 \qquad    0.67  \quad    0.85   \qquad 0.76 ~  \quad      ~     $-$  \quad ~     $-$  \quad ~         $-$ }  \\

\hline\multicolumn{3}{l}{Jafari et al. \cite{Automatic-Detection-of-Melanoma}} & \multicolumn{8}{l}{0.90    \quad    0.72 \qquad    0.70  \quad    0.91   \qquad 0.79 ~  ~      ~     0.79   ~    0.80  ~~         0.61 }  \\

\hline\multicolumn{3}{l}{T. Do et al.  \cite{accessible} Color} & \multicolumn{8}{l}{0.81    \quad 0.73     \qquad 0.66      \quad 0.85       \qquad 0.75 ~  \quad      ~     $-$  \quad ~     $-$  \quad ~         $-$ }  \\

\hline\multicolumn{3}{l}{T. Do et al.  \cite{accessible} Texture} & \multicolumn{8}{l}{0.66    \quad 0.85     \qquad 0.75      \quad 0.79       \qquad 0.78  ~  \quad      ~     $-$  \quad ~     $-$  \quad ~         $-$ }  \\

\hline\multicolumn{3}{l}{T. Do et al.\cite{accessible}Col.and Text.} & \multicolumn{8}{l}{0.84    \quad 0.72     \qquad 0.70      \quad 0.87       \qquad 0.77    ~  \quad      ~     $-$  \quad ~     $-$  \quad ~         $-$ }  \\

\hline\multicolumn{3}{l}{E. Nasr-Esfahani et al. \cite{Nasr}} & \multicolumn{8}{l}{0.81    \quad 0.80      \qquad  0.75    \quad 0.86        \qquad 0.81 ~     \quad         ~     $-$  \quad ~     $-$  \quad ~         $-$}  \\

\hline\multicolumn{3}{l}{\textbf{Our}} & \multicolumn{8}{l}{ \textbf{0.90} ~ \textbf{0.97} ~~~~ \textbf{0.97}  ~ \textbf{0.90}        \qquad \textbf{0.93} ~              ~     \textbf{0.93}   ~     \textbf{0.94}   ~         \textbf{0.87}}  \\
\hline
 &  &                                       &  &  &  &  &  &  &  &                                                                       
\end{tabular}
\label{comparisonMed}
\end{table}



\begin{table}[ht!]
\centering
\caption{Experimental results on Skin-lesion dataset.}
\tiny
\begin{tabular}{lllllllllll}
 &  &                                        &  &  &  &  &  &  &  &                                                                       \\
\hline\multicolumn{3}{l}{Method}                   & \multicolumn{8}{l}{TPR ~~ TNR ~~       PPV ~~    NPV  ~~   ACC   \quad  $F_{l}^{p}$ \quad          $F_{l}^{n}$  \quad         MCC}  \\
\hline\multicolumn{3}{l}{Texture analysis \cite{Textureanalysis}}         & \multicolumn{8}{l}{0.87 ~~~~ 0.71 ~~~        0.76 ~~~ ~-~ ~~~~~~ 0.75~~~~~ -~~ ~~~-~ ~~~ ~~~-}                     \\
\hline\multicolumn{3}{l}{HLIFs \cite{hlifs}}                    & \multicolumn{8}{l}{0.96 ~~~~   0.73 ~~~~~~-  ~~~~~~~- ~~~~~~~ 0.83~~~~~~ -~~ ~~-~  ~~~~~~~- }                             \\
\hline\multicolumn{3}{l}{BIBS \cite{bibis}}                     & \multicolumn{8}{l}{0.92 ~~~~   0.88 ~~~~~     0.91 ~~       -~~~~~~~~  0.90~~~~~~ -~~~ ~-~  ~~~~~~~-}                       \\
\hline\multicolumn{3}{l}{Decision Support \cite{melanomadecision}}         & \multicolumn{8}{l}{0.84 ~~~~  0.79~~~~~~ ~-~  ~ ~ ~-~~~~~~~~~0.81~~~~~~ -~~ ~~-~  ~~~~~~~-}                            \\
\hline\multicolumn{3}{l}{Color pigment boundary \cite{Colorpigmented}}   & \multicolumn{8}{l}{0.95 ~~~~ 0.88 ~~~~      0.92 ~~~  -~~~~~~~~ 0.82~~~~~~ -~~~ ~-~ ~~~ ~~~-} \\
\hline\multicolumn{3}{l}{R. Amelard et al. \cite{Extracting}Asymmetry $F_{C}$ }   & \multicolumn{8}{l}{0.73 ~~~~ 0.64 ~~~~~ ~       -~~~~~  -~~~~~~~~ 0.69~~~~~~ -~~~ ~-~ ~~~ ~~~-}\\
\hline\multicolumn{3}{l}{R. Amelard et al. \cite{Extracting}Proposed HLIFs }   & \multicolumn{8}{l}{0.79 ~~~~ 0.68 ~~~~~ ~       - ~~~~   -~~~~~~~~ 0.75~~~~~~ -~~~ ~-~ ~~~ ~~~-}  \\
\hline\multicolumn{3}{l}{R. Amelard et al. \cite{Extracting}Cavalcanti feature set }   & \multicolumn{8}{l}{0.84 ~~~~ 0.78 ~~~~~ ~       -~  ~~~  -~~~~~~~~ 0.82~~~~~~ -~~~ ~-~ ~~~ ~~~-}\\
\hline\multicolumn{3}{l}{R. Amelard et al. \cite{Extracting}Modified $F_{C}$ }   & \multicolumn{8}{l}{0.86 ~~~~ 0.75 ~~~~~ ~       -~  ~~~  -~~~~~~~~ 0.72~~~~~~ -~~~ ~-~ ~~~ ~~~-}\\
\hline\multicolumn{3}{l}{R. Amelard et al. \cite{Extracting}Combined $F_{MC}$ $F_{A}^{HLIFS}$ }   & \multicolumn{8}{l}{0.91 ~~~~ 0.80 ~~~~~ ~       -~  ~~~  -~~~~~~~~ 0.86~~~~~~ -~~~ ~-~ ~~~ ~~~-}\\
\hline\multicolumn{3}{l}{\textbf{Our}} & \multicolumn{8}{l}{ \textbf{0.84} ~~~ \textbf{0.92} ~~~ \textbf{0.91}  ~ \textbf{0.85}  ~~~ \textbf{0.88} ~                  \textbf{0.87}   ~     \textbf{0.88}   ~         \textbf{0.76}}  \\

\hline
 &  &                                        &  &  &  &  &  &  &  &                                                                      
\end{tabular}
\label{comparisonSkin}
\end{table}

\section{Conclusions and Future Works}
\label{conc}

The challenge in the discrimination of melanoma and nevi has resulted to be very interesting in recent years. The complexity of the task is linked to different factors such as the large amount of types of melanomas or the difficulties for digital phase acquisition (noise, lighting, angle, distance and much more). Machine learning classifiers suffer greatly these factors and inevitably reflect on the quality of the results. In support, the convolutional neural networks give a big hand for both classification and features extraction phases. In this context, we have proposed a framework that combines standard classifiers and features extracted with convolutional neural networks using a transfer learning approach. The results produced certainly support the theoretical thesis. A multiple representation of the image compared to a single one is a high discrimination factor even if the features adopted are completely abstract. The extensive experimental phase has shown how the proposed approach is competitive, and in some cases surpassing, with respect to state of the art methods. Certainly, the main weak point concerns the computational complexity relating to features extraction phase, as it is known, takes a long time especially when the data to be processed grows. Future work will certainly concern the study and analysis of additional convolutional neural networks still unexplored for this type of problem or, alternatively, the application of the proposed framework to tasks different from the melanoma detection.

\section*{Acknowledgements}

Our thinking is for Alfredo Petrosino. He followed us during first steps towards the Computer Science, through a whirlwind of goals, ideas and, specially, love and passion for the work. We will be forever grateful great master.


\bibliography{bibliography}

\end{document}